%Paper: hep-th/9407194
%From: abdalla@surya11.cern.ch (Elcio Abdalla)
%Date: Fri, 29 Jul 94 15:48:21 +0200

\magnification 1200
%\magnification = 1200
%

%
\font\eightrm=cmr8
\font\eighti=cmmi8
\font\eightsy=cmsy8
\font\eightbf=cmbx8
\font\eighttt=cmtt8
\font\eightit=cmti8
\font\eightsl=cmsl8
\font\sixrm=cmr6
\font\sixi=cmmi6
\font\sixsy=cmsy6
\font\sixbf=cmbx6
\catcode`@11
\newskip\ttglue
\font\grrm=cmbx10 scaled 1200

\def\eightpoint{\def\rm{\fam0\eightrm}
\textfont0=\eightrm \scriptfont0=\sixrm \scriptscriptfont0=\fiverm
\textfont1=\eighti \scriptfont1=\sixi \scriptscriptfont1=\fivei
\textfont2=\eightsy \scriptfont2=\sixsy \scriptscriptfont2=\fivesy
\textfont3=\tenex \scriptfont3=\tenex \scriptscriptfont3=\tenex
\textfont\itfam=\eightit \def\it{\fam\itfam\eightit}
\textfont\slfam=\eightsl \def\sl{\fam\slfam\eightsl}
\textfont\ttfam=\eighttt \def\tt{\fam\ttfam\eighttt}
\textfont\bffam=\eightbf
\scriptfont\bffam=\sixbf
\scriptscriptfont\bffam=\fivebf \def\bf{\fam\bffam\eightbf}
\tt \ttglue=.5em plus.25em minus.15em
\normalbaselineskip=6pt
\setbox\strutbox=\hbox{\vrule height7pt width0pt depth2pt}
\let\sc=\sixrm \let\big=\eightbig \normalbaselines\rm}
\newinsert\footins
\def\newfoot#1{\let\@sf\empty
  \ifhmode\edef\@sf{\spacefactor\the\spacefactor}\fi
  #1\@sf\vfootnote{#1}}
\def\vfootnote#1{\insert\footins\bgroup\eightpoint
  \interlinepenalty\interfootnotelinepenalty
  \splittopskip\ht\strutbox % top baseline for broken footnotes
  \splitmaxdepth\dp\strutbox \floatingpenalty\@MM
  \leftskip\z@skip \rightskip\z@skip
  \textindent{#1}\footstrut\futurelet\next\fo@t}
\def\fo@t{\ifcat\bgroup\noexpand\next \let\next\f@@t
  \else\let\next\f@t\fi \next}
\def\f@@t{\bgroup\aftergroup\@foot\let\next}
\def\f@t#1{#1\@foot}
\def\@foot{\strut\egroup}
\def\footstrut{\vbox to\splittopskip{}}
\skip\footins=\bigskipamount % space added when footnote is present
\count\footins=1000 % footnote magnification factor (1 to 1)
\dimen\footins=8in % maximum footnotes per page

\def\ref#1{$^{#1}$}
\def\flex{\raise 6pt\hbox{$\leftrightarrow $}\! \! \! \! \! \! }
\def\oversome#1{ \raise 8pt\hbox{$\scriptscriptstyle #1$}\! \! \! \! \! \! }
\def\tr{ \mathop{\rm tr}}

\newbox\bigstrutbox
\setbox\bigstrutbox=\hbox{\vrule height10pt depth5pt width0pt}
\def\bigstrut{\relax\ifmmode\copy\bigstrutbox\else\unhcopy\bigstrutbox\fi}
\def\refer[#1/#2]{ \item{#1} {{#2}} }
\def\rev<#1/#2/#3/#4>{{\it #1\/} {\bf#2}, {#3}({#4})}
\def\boxit#1{\vbox{\hrule\hbox{\vrule\kern3pt
\vbox{\kern3pt#1\kern3pt}\kern3pt\vrule}\hrule}}

\def\2figure#1#2#3#4{\vbox{ \hrule width#1truecm \hbox{\vrule height#2truecm
\hskip #1truecm
\vrule height#2truecm }\hrule width#1truecm \hbox{\vrule\vbox{\hsize #1truecm
\baselineskip=10pt
\noindent\strut#3}\vrule}\hrule width#1truecm
\hbox{\vrule\vbox{\hsize #1truecm
\baselineskip=10pt
\noindent\strut#4}\vrule}\hrule width#1truecm  }}
\def\3figure#1#2#3#4#5{\vbox{ \hrule width#1truecm \hbox{\vrule height#2truecm
\hskip #1truecm
\vrule height#2truecm }\hrule width#1truecm \hbox{\vrule\vbox{\hsize #1truecm
\baselineskip=10pt
\noindent\strut#3}\vrule}\hrule width#1truecm
 \hbox{\vrule\vbox{\hsize #1truecm
\baselineskip=10pt
\noindent\strut#4}\vrule}
\hrule width#1truecm \hbox{\vrule\vbox{\hsize #1truecm
\baselineskip=10pt
\noindent\strut#5}\vrule}\hrule width#1truecm  }}

\def\sqr#1#2{{\vcenter{\hrule height.#2pt
   \hbox{\vrule width.#2pt height#1pt \kern#1pt
    \vrule width.#2pt}
    \hrule height.#2pt}}}

% Here are my additional definitions:

\def\smin{\,\raise 0.06em \hbox{${\scriptstyle \in}$}\,}
\def\smsubset{\,\raise 0.06em \hbox{${\scriptstyle \subset}$}\,}

\def\Natural{\hbox{\hskip 1.5pt\hbox to 0pt{\hskip -2pt I\hss}N}}

\def\Rational{\hbox{\hbox to 0pt{\hskip 2.7pt \vrule height 6.5pt
                                  depth -0.2pt width 0.8pt \hss}Q}}
\def\Real{\hbox{\hskip 1.5pt\hbox to 0pt{\hskip -2pt I\hss}R}}
\def\Complex{\hbox{\hbox to 0pt{\hskip 2.7pt \vrule height 6.5pt
                                  depth -0.2pt width 0.8pt \hss}C}}
%%%%%%%%%%%%%%%%%
% definitions for the second book
%tcap1

\def \E {{{\rm e}}}

%%%%%%%%tcap2

%%%%%%%%tcap3
\def \1ok{{1\over \kappa ^2} }

\def \3dslim {{\rm DS}\!\!\!\!\!\!\!\!\lim }
\def \4dslim {{\rm DS}\!\!\!\!\!\!\!\!\!\!\lim }
\def \tr {{\rm tr}\, }

\def \2kk{\left( \matrix {2k\cr k\cr }\right) }
\def \Rs4{{R^k\over 4^k} }

%%%%%%%%%tcap4
\def \1ok{{1\over \kappa ^2} }

%%%%%%%%%%tcap5

%%%%%%%%%%tcap6
%\def \DSLim {{\rm DS}\!\!\lim }
%%%%%%%%%%tcap7
%nada
%%%%%%%%%%tcap8
%nada
%%%%%%%%%appb

\def \E {{\rm e}}
\nopagenumbers
\hfill CERN-TH.7365/94

\hfill hep-th/9407194
\vskip .3cm
\centerline{\grrm Higher algebras and mesonic spectrum in two-dimensional QCD }
\vskip 1.5cm
\centerline {E. Abdalla\newfoot {${}^*$}{Permanent address: Instituto de
F\'\i sica - USP, C.P. 20516, S. Paulo, Brazil.} and  M.C.B. Abdalla\newfoot
{${}^{**} $}{Permanent address: Instituto de F\'\i sica Te\'orica - UNESP,
R. Pamplona 145, 01405-000, S.Paulo, Brazil.}, }
\vskip 1.5truecm
\centerline { CERN, Theory Division, CH-1211 Geneva 23, Switzerland}
\vskip 2cm

\centerline{\bf Abstract}
\vskip .5cm
\noindent We construct composite operators in two-dimensional bosonized QCD,
which obey a $W_\infty$ algebra, and discuss their relation to analogous
objects recently obtained in the fermionic language. A complex algebraic
structure is unravelled, supporting the idea that the model is integrable.
For singlets we find a mass spectrum obeying the Regge behavior.
\vfill

\noindent CERN-TH.7365/94

\noindent July 1994

\noindent hep-th/9407194
\vskip 1cm
\eject
\countdef\pageno=0 \pageno=1
\newtoks\footline \footline={\hss\tenrm\folio\hss}
\def\folio{\ifnum\pageno<0 \romannumeral-\pageno \else\number\pageno \fi}
\def\advancepageno{\ifnum\pageno<0 \global\advance\pageno by -1
\else\global\advance\pageno by 1 \fi}

Two-dimensional gauge theories have always offered a good laboratory of ideas
in Quantum Field Theory\ref{1}. Quantum chromodynamics (QCD$_2$), in
particular, is an extraordinarily difficult problem, whose solution in terms of
string ideas\ref{2} has not been fully accomplished. Moreover, in spite of the
fact that Abelian gauge theories are soluble (see [1] and references therein),
the non-Abelian case has evaded a full solution, although a number of authors
achieved considerable progress\ref{1-8}. Further understanding has been
obtained in [9], where it has been proved that fermion bilocal operators obey
a $W_\infty$-type algebra, opening a possibility of obtaining the full mesonic
spectrum of the theory. In particular, $1/N$ corrections turn out to be
feasible in such a scheme.

In a recent paper\ref{10}, we have rewritten QCD$_2$ in terms of bosonic
fields by integrating out the fermions, obtaining an integrable theory. Indeed,
we start out of the generating functional
$$
{\cal Z}[\eta, \overline \eta, i_\mu] = \int {\cal D}\psi {\cal D}\overline
\psi {\cal D}A_\mu \, \E^{i \int {\rm d}^2x \, \left[ - {1\over 4} \tr F_
{\mu\nu} F^{\mu\nu} + i \overline \psi \not D \psi + \overline \eta \psi  +
\overline \psi \eta + i_\mu A^\mu \right]}\quad ,\eqno(1)
$$
with $D_\mu = \partial _\mu - ieA_\mu$, $F_{\mu\nu} = \partial _\mu A_\nu -
\partial _\nu A_\mu - ie [A_\mu ,A_\nu]$, $\eta$ and $\overline \eta$ are the
fermionic sources and $i_\mu$ is the gauge source. If we make the change of
variables
$$
A_+= {i\over e} U^{-1}\partial _+U\quad ,\quad A_-= {i\over e} V\partial _-
V^{-1}\quad , \eqno(2)
$$
it is possible to compute the fermion determinant in terms of the WZW
action\ref{4} as
$$
\det i \not \!\! D  = \E^{i\Gamma[UV]} \det i\not \! \partial\quad ,\eqno(3)
$$
with
$$
\Gamma[g] = {1\over 8\pi}\int {\rm d}^2 x\, \partial ^\mu g^{-1} \partial _\mu
g
+ {1\over 4\pi} \epsilon ^{\mu\nu} \int {\rm d} r \int {\rm d}^2x\,\hat g
^{-1} \dot {\hat g} \hat g^{-1} \partial _\mu \hat g
\hat g^{-1} \partial _\nu \hat g \quad ,\eqno(4)
$$
which obeys the Polyakov--Wiegman identity\ref{4}
$$
\Gamma[UV] = \Gamma[U] + \Gamma[V]+{1\over 4\pi}\tr\int{\rm d}^2x \,U^{-1}
\partial_+ U \,V\partial_- V^{-1}\quad .\eqno(5)
$$

The fully bosonized form of QCD$_2$ is obtained using the invariance of the
Haar measure, writing the fermionic determinant as
$$
\det i \not \!\! D = \int {\cal D} \, g \, \E^{i S_F[A,g]} \quad ,\eqno(6)
$$
where
$$
\eqalign{
S_F [A,g] & = \Gamma[UgV] - \Gamma[UV] \cr
& = \Gamma [g] + {1\over 4\pi} \int {\rm d}^2x\, \left[e^2 A_\mu A_\mu-e^2 A g
\overline A g^{-1} - ie A g \overline \partial g^{-1} - ie \overline A g^{-1}
\partial g \right] \quad .\cr}\eqno(7)
$$

After some algebraic manipulations and definning $\widetilde g = U g V$ we
arrive at the result
$$
\eqalign{
&{\cal Z}\left[ \overline \eta , \eta , i_\mu\right]= \int {\cal D}
\widetilde g \, \E^{i \Gamma[\tilde g]}\int  {\cal D}E {\cal D}U {\cal D}V
{\cal D} ({\rm ghosts})\times \cr
& \times \E^{-i(c_V+1)\Gamma[UV] -i \int{\rm d}^2x\, \tr[{1\over 2} E^2 -
{1\over 2}EF_{+-}]+iS_{{\rm ghosts}}+i\int {\rm d}^2 x\, i_\mu A_\mu -i
\int {\rm d}^2x {\rm d}^2y \, \overline \eta  (x) (i \not D)^{-1}(x,y)
\eta  (y) }\quad .\cr}\eqno(8)
$$
In order to obtain the above result, notice that the $E$-integration is
Gaussian, reproducing the gauge field strenght squared, and the Casimir $c_V$
is a consequence of the  change of variables ${\cal D}A_+ {\cal D}A_- ={\cal D}
U {\cal D}V \, \E^{i c_V \Gamma[UV]}$ (or else the Dirac operator determinant
(3) in the adjoint representation).

We can substitute the gauge field in terms of eq. (2) and try to rewrite the
theory using the gauge-invariant combination $\Sigma = UV$, such as in
$$
\tr E F_{+-}= {i\over e} \tr UEU^{-1} \partial_+ (\Sigma \partial_-\Sigma
^{-1}) \quad .\eqno(9)
$$

At this point we change variables as
$$
U E U^{-1} = 2ie (c_V +1) {1\over 4\pi} \partial_+^{-1}(\beta^{-1}\partial_+
\beta) \quad ,\eqno(10)
$$
and obtain the final generating functional
$$
\eqalign{
&{\cal Z}\left[ \overline \eta , \eta , i_\mu\right]=\int{\cal D}\widetilde g
\, \E^{i\Gamma[\tilde g]}{\cal D} ({\rm ghosts})\, \E^{i S_{{\rm ghosts}}}
\int {\cal D} \widetilde
\Sigma\,\E^{-i(c_V+1)\Gamma[\widetilde\Sigma]}\times\cr
&\times \int {\cal D} \beta\, \E^{i\Gamma[\beta] +i {\lambda^2\over 2}\tr \int
{\rm d}^2 x\,[\partial_+^{-1}(\beta^{-1}\partial_+\beta)]^2}\,\E^{i\int
{\rm d}^2x\, i_\mu A_\mu -i\int{\rm d}^2x\,{\rm d}^2y\overline \eta (x)
(i \not D)^{-1}(x,y) \eta (y) }\quad ,\cr}\eqno(11)
$$
where $\lambda = {c_V+1\over 2\pi}e$. Here we distinguish three apparently
independent sectors, namely the (conformally invariant) $\widetilde \Sigma$
and $\widetilde g$ sectors, and the off-critically perturbed $\beta$
sector. As discussed in ref. [10] (see also [12]) such sectors interact via the
constraints arising from the BRST structure of the theory. The $\beta$-sector
turns out to be integrable. Indeed, the $\beta$-equation of motion can be
written in terms of
$$
\eqalignno{
J^\beta_+&= \beta^{-1}\partial_+ \beta\quad ,& (12a)\cr
J^\beta_-&= 4\pi \lambda^2 \partial_+^{-2}J_+ \quad ,& (12b)\cr}
$$
as
$$
[D_+, D_-] = [\partial _+ - J^\beta_+,\partial _--J^\beta_-]=0\quad ,\eqno(12c)
$$
or else as a non-linear conservation law
$$
\partial _+I_-^\beta = \partial _+ \, \left\{  4\pi \lambda^2 J^\beta_- -
\partial _+\partial _- J^\beta_- + [J^\beta_-,\partial _+J^\beta_-] \right\} =
0 \quad .\eqno(13)
$$

A duality-type transformation can be made in (11) in order to rewrite it in
terms of fields appropriately describing the strong coupling (low-energy)
limit. One uses
$$
\E^{{i\over 2}\lambda^2\int{\rm d}^2x\,\left[\partial_+^{-1}(\beta^{-1}
\partial_+\beta)\right]^2}=\int{\cal D}C_-\,\E^{i\int{\rm d}^2 x\,\left[ {1
\over 2}(\partial_+C_-)^2+\lambda\tr C_-\beta^{-1}\partial_+\beta\right]}\quad
,
\eqno(14)
$$
and changes variables according to $C_-={1\over 4\pi\lambda}W\partial_-W^{-1}$,
arriving at
$$
{\cal Z} = \int {\cal D} \widetilde \beta \,\E^{i\Gamma[\tilde \beta]}\int
{\cal D}W\,\E^{-i(c_V+1)\Gamma[W]-{i\over 2(4\pi\lambda)^2}\tr \int{\rm d}^2 x
\, [\partial(W\overline \partial W^{-1})]^2}\quad ,\eqno(15)
$$
where $\widetilde \beta = \beta W$. The $W$-equation of motion corresponds to
an integrability condition similar to the $\beta$-formulation, and can be
written as a conservation law in the form
$$
\partial_+ I^W_-=\partial _+ \left\{ {1\over 4\pi}(c_V+1)J_-^W - {1\over (4\pi
\lambda)^2}\partial_+\partial_-\,\overline J_-^W - {1\over (4\pi\lambda)^2}
[J_-^W, \partial_+ J_-^W]\right\}=0\quad ,\eqno(16)
$$
where $J_-^W= W \partial _-W^{-1}$.

In terms of (dual-equivalent)\ref{11} theories (11) and (15), it is difficult
to obtain the current algebra, due to the complicated interaction terms, namely
the non-local $\beta$-interaction on the one hand, and the higher derivative
$W$-interaction on the other hand. Therefore we introduce auxiliary fields,
rendering in both formulations acceptable results, in the sense that it is
possible to perform the canonical procedure. In such a case we have the
$\beta/W$ effective actions
$$
\eqalignno{
S[\beta] & = \Gamma[\beta] + {1\over 2}\int {\rm d}^2 x \,\tr (\partial
_+C_-)^2
+ \lambda \int {\rm d}^2 x \,\tr C_-\beta^{-1}\partial _+\beta \quad ,&(17a)\cr
S[W]     & = - (c_V+1) \Gamma[W] + {1\over 2} \tr \int {\rm d}^2x \,
\left[ - B^2 + {1\over 2\pi \lambda} \partial _+B \partial _-W W^{-1} \right]
\quad . &(17b)\cr}
$$

We note here the minus sign in front of the WZW term in the $W$-formulation
(17$b$), signalling the presence of negative metric states. However this is
not the full story. As a matter of fact, the complete system is described by
the
partition function (11/15), which as discussed in [10], based on the
Karabali-Schnitzer argumentation\ref{12}, presents several constraints. The
first-class constraints\ref{12} select the Hilbert space, defining the
appropriate cohomology, representing a GKO construction\ref{13}. The theory
thus defined has positive metric. In the present case there are also
second-class constraints\ref{10}. These are more complicated, and one is
obliged to introduce the Dirac\ref{14} formulation in order to find the
commutative algebras.

We start with the Poisson algebra. Using the formulation of ref. [6] we find
(see also [10])
$$
\eqalignno{
\Pi_- & = \partial _+C_-\quad , &(18a)\cr
\widetilde {\hat \Pi}^\beta & = {1\over 4\pi} \partial _0 \beta^{-1} + \lambda
C_- \beta^{-1} \quad , &(18b)\cr}
$$
where $\; \widetilde{}\; $ means transposition with respect to the group
indices, and $\widetilde {\hat \Pi}^\beta$ is the local part of the $\beta$
canonical momentum (i.e. neglecting the WZW term), and satisfies
$$
\eqalignno{
\{ \beta_{ij}(t,x) , \widetilde {\hat \Pi}^\beta_{kl} (t,y) \} & = \delta_{il}
\delta_{kj} \delta (x-y) \quad , & (19a)\cr
\left\{ \widetilde{\hat \Pi}^\beta_{ji}(t,x), \widetilde{\hat \Pi}^\beta_{lk}
(t,y)\right\} & =-{1\over 4\pi}\left( \partial_1\beta^{{}^{-1}}_{_{jk}}
\beta^{{}^{-1}}_{_{li}}-\partial_1\beta^{{}^{-1}}_{_{li}}
\beta^{{}^{-1}}_{_{jk}}\right)\delta(x-y)\quad . &(19b)\cr}
$$

The $C_-$ equation of motion leads to its definition
$$
C_- = {1\over 4\pi \lambda } J_-^\beta = \lambda \partial _+^{-2}
(\beta^{-1}\partial _+\beta)\quad .\eqno(20)
$$

On the other hand, for the $W$-formulation (dual) we have
$$
\eqalignno{
\Pi^W_{ij} &= {\partial S\over \partial \partial _0W_{ij}} = -{1\over 4\pi}
(c_V+1) \partial _0W^{-1}_{ji} - {1\over 4\pi} (c_V+1)A_{ji} + {1\over 4\pi
\lambda}(W^{-1} \partial _+B)_{ji}\, ,&(21a)\cr
& = \hat \Pi^W_{ij} - {1\over 4\pi} (c_V +1) A_{ji}\quad ,&(21b)\cr}
$$
where $A_{ij}$ is the contribution from the topological term to the momentum.
There is no local representation for $A_{ij}$, but only its derivatives are
necessary, i.e.
$$
F_{ij;kl} = {\delta A_{ij}\over \delta W_{lk}} - {\delta A_{kl}\over
\delta W_{ji}} = \partial _1 W^{-1}_{il} W^{-1}_{kj} - W^{-1} _{il} \partial
_1 W^{-1}_{kj}\quad .\eqno(21c)
$$

The current and its $+, -$ derivatives are, in phase space, given by
$$
\eqalignno{
J^W_- & = W\partial _-W^{-1} = - 4\pi \lambda \widetilde \Pi_B\quad ,&(22a)\cr
\partial_+J^W_-&=-4\pi\lambda\partial_+\widetilde \Pi_B=4\pi\lambda B\quad . &
(22b)\cr}
$$

The second-class constraints are obtained by coupling a subset of fields to an
external gauge field $A_+^{^{ext}}$ for the $\beta$-formulation or
$A_-^{^{ext}}$ for the $W$-formulation. One thus obtains two self-commuting
constraints, but their difference is second-class, leading to the constraint
$$
\Omega^\beta_{ij}=(\beta \partial _-\beta^{-1})_{ij} + 4\pi \lambda (\beta C_-
\beta^{-1})_{ij} - (\widetilde g \partial _-\widetilde g^{-1})_{ij}\sim 0
\quad ,\eqno(23a)
$$
or
$$
\Omega ^W = (c_V+1) \Sigma^{-1}\partial _+ \Sigma - (c_V + 1) W^{-1} \partial_+
W + {1\over \lambda} W^{-1} \partial _+ B W \sim 0 \quad , \eqno(23b)
$$
which in phase space do not depend on the auxiliary field, i.e.
$$
\eqalignno{
\Omega_{ij}^\beta & =4\pi (\beta\widetilde{\hat\Pi}^\beta)_{ij}+\partial_1\beta
\beta^{-1} - 4\pi (\widetilde g \widetilde {\hat \Pi}^{\tilde g})_{ij}-\partial
_1\widetilde g \widetilde g^{-1}\sim 0\quad , &(24a)\cr
\Omega ^{W} &= - \widetilde {\hat \Pi}^WW + {1\over 4\pi} W^{-1} \partial_1 W +
\widetilde{\hat\Pi}^\Sigma\Sigma-{1\over 4\pi}\Sigma^{-1}\partial_1\Sigma\sim 0
\quad . &(24b)\cr}
$$

Only the currents
$$
\eqalignno{
j_-^\beta & = 4\pi \beta \widetilde {\hat \Pi}^\beta + \partial_1 \beta \beta
^{-1} \quad ,& (25a)\cr
j_+^W & = 4\pi \widetilde {\hat \Pi} W - W^{-1} \partial _1 W \quad ,
&(25b)\cr}
$$
enter in the constraint, and they commute with the auxiliary fields, resp. $(
C_-, \Pi_-\!), (\!B, \Pi_B\!)$, as well as with the currents $J_-^B\, ,\,
J_-^W\, ,\, j_+^\beta $ and $ j_+^W$. Therefore, for the latter objects Poisson
and Dirac structures are the same. In particular
$$
\left[ {J_-^\beta }_{ij}(x), {\partial _+J_-^\beta}_{kl} (y)\right] = (4\pi
\lambda)^2 i \delta_{kj}\delta_{il}\delta(x^1-y^1)\quad ,\eqno(26a)
$$
and
$$
\left[ {J_-^W }_{ij}(x), {\partial _+J_-^W}_{kl} (y)\right] = (4\pi
\lambda)^2 i \delta_{kj}\delta_{il}\delta(x^1-y^1)\quad .\eqno(26b)
$$

Observe the duality of the phase space, where the role of $(C_-, \Pi_-)$ is
interchanged (in inverse order!) with that of $(\Pi_B, -B)$.

We define $J(x) = {1\over 4\pi \lambda} J_-(x)$, for both ($J_-^{\beta, W}$),
the bilocal
$$
M(x,y) = \tr \, J(t,x) \partial _+ J(t,y)\quad ,\eqno(27)
$$
and we are led to the $W_\infty$-algebra
$$
\left[M(x,y),M(z,w)\right]=i\delta(x-w)M(z,y)-i\delta(z-y)M(x,w)\quad.\eqno(28)
$$

Time evolution may be obtained from the Hamiltonian, which in the
$\beta$-formulation reads
$$
H_\beta = -{1\over 16\pi} \left[ \left( J_+^{\beta}\right)^2 + \left( j_-^
{\beta}\right)^2 \right] - {1\over 2} \lambda J^\beta_+ C_- +
\pi \lambda^2 C_-^2 + {1\over 2} \Pi\left(\Pi_- - 2C'\right)\quad .\eqno(29)
$$

We still have to take into account the constraint, which eliminates $j_-$ in
terms of the $\widetilde g$ fields. This procedure is safe, once $j_-$ commutes
with the variables considered before, namely $C_-\, ,\, \Pi_-\, ,\, J_+$ and
$J_-$. The commutators of the Hamiltonian with $M(x,y)$ leads to
$$
-i[H, M(x,y)] = \left( {\partial \over \partial x} + {\partial\over\partial y}
\right) M(x,y;t) - P(x,y;t) - \lambda Q^{(1)} (x,y;t)\quad ,\eqno(30)
$$
where
$$
\eqalignno{
P(x,y;t) & = \tr \partial _+ \, J(t,x) \partial _+ J(t,y)\quad , &(31a)\cr
Q^{(1)}(x,y;t) & = \tr \, J(t,x) \partial^2_+ J(t,y)\quad .&(31b)\cr}
$$

In terms of the phase-space variables, we have
$$
\eqalignno{
P(x,y;t) & = \tr\, \Pi (t,x) \Pi(t,y)\quad , &(32a)\cr
Q^{(1)}(x,y;t) & = \tr \, C_-(t,x) J_+(t,y)\quad ,&(32b)\cr}
$$
where
$$
J_+(t,x) = j_+(t,x) + 4\pi \lambda C_-(t,x)\quad .\eqno(33)
$$

If we try to include such fields in the $M(x,y)$ algebra, we generate an
infinite number of terms. We computed
$$
-i[P(x,y), M(z,w)]  = -P(x,w)\delta(y-z) - P(y,w)\delta(x-z)\quad,\eqno(34)
$$
which closes, but
$$
-i[Q^{(1)}(x,y;t), Q^{(1)}(z,w;t)]  = Q^{(2)} (x,z;y;t)\delta(y-w) - Q^{(2)}
(z,x;y;t)\delta(y-w)\quad,\eqno(35)
$$
up to contact terms, where $Q^{(2)}(x,z;y;t)= \tr C_-(x)C_-(z)J_+(y)$.
Moreover we find also
$$
\eqalignno{
-i[P(x,y),Q^{(1)}(z,w)]= & -4\pi\lambda\left[ M(z,y)\delta(x-w)+M(z,x)
\delta(y-w)\right] - \cr
& - R(y,w)\delta (x-z) - R(x,w)\delta (y-z)\quad ,&(36)\cr}
$$
where $R(x,y) = \tr \Pi(x)J_+(y)$. In fact, here, there are infinite new terms,
which get generated by multiple commutators, such as $Q^{(n)} (\{ z_i\} , w) =
\tr C_-(z_1) \cdots C_-(z_n) J_+(w)$, as well as anlogous $R^{(n)}(\{ z_i\} ,
w)$ terms.

Such $W_\infty$-type algebras appear not only in QCD$_2\,$\ref{9} but also in
the description of the incompressible fluid of the Quantum Hall effect\ref{15},
where however only singlets are considered.

If we sum over all symmetry group indices, i.e. consider only singlets, a
further simplified structure arises. Taking the trace of (13) [or equivalently
(16)], we have the conserved charges
$$
\eqalignno{
Q^{\beta} &= \tr \int {\rm d}x^1\left\{ 4\pi\lambda^2J_-^\beta - \partial
_+\partial _-J_-^\beta\right\}\quad ,&(37a)\cr
Q^{W} &= \tr \int {\rm d}x^1\left\{ 4\pi\lambda^2 (c_V+1)J_-^W - \partial
_+\partial _-J_-^W\right\}\quad .&(37b)\cr}
$$

We can use the phase-space formulation for
$$
\eqalignno{
J_-^\beta &= 4\pi \lambda C_-\quad ,&(38a)\cr
\partial _+\partial _-J_-^\beta &= 4\pi \lambda ^2 j^\beta_+ + (4\pi\lambda)^2
\lambda C_-\quad ,&(38b)\cr}
$$
obtaining
$$
Q^\beta = -4\pi \lambda^2 \int {\rm d}x^1 \, \tr j_+^\beta\quad ,\eqno(39)
$$
with $j_+^\beta = -4\pi \widetilde {\hat \Pi}^\beta \beta + \beta ^{-1}\beta'$.

The algebraic structure just discussed is not the only higher algebra
underlying the theory. As we discussed in a previous publication, the currents
$I_-(x^-)$ obey an affine Lie structure given by\ref{16}
$$
\left\{ I_-^{ij} (x), I_-^{kl} (y)\right\} = \left( \delta^{il} \delta^{kj} -
\delta^{kj}\delta^{il}\right) \delta(x-y) + {c_V+1\over 2\pi} \delta_{il}
\delta_{kj} \delta'(x-y)\quad ,\eqno(40)
$$
and $\widehat J_-(x^+,x^-)$ is a realization of such algebra,
$$
\left\{ I_-^{ij} (x), \widehat J_-^{kl} (y^+,y^-)\right\} = \left( \delta^{il}
\widehat J_-^{kj} - \delta^{kj} \widehat J_-^{il}\right) \delta(x^--y^-) +
\delta_{il} \delta_{kj} \delta'(x^--y^-)\quad ,\eqno(41)
$$
where $\widehat J_-$ is either $J_-^W$ or $J_-^\beta$ with a suitable
normalization factor. Therefore, $\partial _+\widehat J_-$ is a primary field
of the $I_-$ affine algebra\ref{16,17}, depending on $x^+$ as a parameter.

We should however stress the fact that while $I_-$ is a {\it conserved} current
$(\partial _+I_-=0)$, the other higher operators are not. Therefore their
action generates new states of the theory.

The presence of higher conservation laws, and the complex algebraic structure
is characteristic of integrable systems\ref{18-20}, confirming recent claims
that two-dimensional QCD\ref{20,10}, or else high-energy scattering of
strongly interacting systems\ref{21} are described by integrable quantum field
theories\ref{22,23}.

The problem simplifies drastically upon consideration of singlets only. Indeed,
as discussed above, $\tr j_+$ is a right-moving field. The $(++)$ component of
the energy momentum tensor is given by
$$
\eqalignno{
T_{++} = \, & \, {1\over N+1}\left\{-{1\over 16\pi} (J_+)^2+{1\over 2}
(\partial
_+ C_-)^2\right\}\cr
= \, & \, {1\over N+1}\left\{ - {1\over 16\pi} (j_+)^2 - {1\over 2}\lambda j_+
C_- - \pi \lambda^2 C_-^2 + {1\over 2} (\partial _+ C_-)^2\right\}\quad ,&(42)
\cr }
$$
where we introduced the factor ${1\over N+1}$ in order for the limit $N \to
\infty$ to be well defined, in accordance with the Sugawara
construction\ref{16}.

Consideration of the trace part leads to a left-moving $C_-$ field and the last
term above drops off. Therefore we are left with the mode expansions
$$
\eqalignno{
j_+^{\tr} & = i \lambda \sum j_n \, \E^{in\lambda (x-t)}\quad,&(43a)\cr
C_-^{tr}  & = i \sum C_n \, \E^{in \lambda (x-t)}\quad ,&(43b)\cr}
$$
with the following commutation relation for the modes
$$
\eqalignno{
[j_n ,j_m ] & = 4n \delta_{n, -m}\quad ,&(44a)\cr
[C_n , C_m] & = {1\over 2\pi n} \delta _{n,-m}\quad .&(44b)\cr}
$$

The integral of (42) is given by
$$
\eqalignno{
T_{++}^{(0)} = \, & \, {\lambda^2\over 4(N+1)} \sum _{n\ge 1} j_{-n}j_{n} +
{\lambda^2\over 8(N+1)} j_0^2  + {\pi\over (N+1)}\lambda^2\sum \left( j_{-n}
C_n + C_{-n}j_n\right)\cr
\, & \, + {4\pi^2\lambda^2\over N+1}\sum C_{-n}C_n + {2\pi^2\lambda^2\over
N+1}C_0^2\quad .&(45)\cr}
$$

The zero-mode component can be interpreted as mass squared up to an
undetermined constant, and
$$
{\lambda^2\over N+1} = {e^2(N+1)\over (2\pi)^2} = {e^2_{fin}\over (2\pi)^2}
\quad ,\eqno(46)
$$
where $e_{fin}$ is defined as a new coupling constant\ref{3} in the limit $N\to
\infty$. We thus find
$$
\eqalignno{
T^{(0)} j_{-n}\vert 0 \rangle = & {\lambda^2\over N+1} \left(nj_{-n} + 4n \pi
C_{-n}\right)\vert 0 \rangle\quad ,&(47a)\cr
T^{(0)} C_{-n}\vert 0 \rangle = & {\lambda^2\over N+1} \left( {1\over 2 n}
j_{-n} + {2\pi\over n}C_{-n} \right)\vert 0 \rangle\quad .&(47b)\cr}
$$

The eigenvectors are of the form $j_{-n} + \xi C_{-n}$, and we obtain
$$
\xi_\pm = - n\left( n - {2\pi\over n}\right) \left[ 1 \pm \sqrt{1+ {8\pi\over
\left(n - {2\pi\over n}\right)^2}}\right]\quad .\eqno(48)
$$

For large values of $n$, we have
$$
\eqalignno{
\xi_+ = - & 2 (n^2 - 2\pi)\quad ,&(49a)\cr
\xi_- = & 4\pi\quad ,&(49b)\cr}
$$
for which we find, in the first case, mass eigenvalues of the type
$$
m^2 \sim e^2_{fin}\times {\cal O} \left( {1\over n^5} \right)\quad ,\eqno(50a)
$$
or
$$
m^2 \sim n \, e^2_{fin} \quad ,\eqno(50b)
$$
therefore, the former is degenerate, while the latter presents a Regge
behaviour, in accordance with ref. [3]. Thus it is possible in principle
to find 't Hooft's results\ref{3}, as well as corrections to it. Relations with
collective fields are yet to be discovered\ref{24}.
\vskip 1cm

\noindent Acknowledgements: The authors would like to thank Luiz
Alvarez-Gaum\'e and Guillermo Zemba for discussions. This work was partially
supported by CAPES (E.A.), Brazil, under contract No.1526/93-4, and by CNPq
(M.C.B.A.), Brazil, under contract No.204220/77-7.
\vskip 1cm
\penalty-3500

\centerline {\bf References}
\vskip .5cm
\nobreak
\refer[[1]/E. Abdalla, M.C.B. Abdalla and K. Rothe, {``Non-perturbative
methods in two-\-dimen\-sional Quantum Field Theory"}, World Scientific, 1991]

\refer[[2]/D. Gross, Nucl. Phys. {\bf B400} (1993) 161;]

\refer[/D. Gross and W. Taylor, Nucl. Phys. {\bf B400} (1993) 181; {\bf B403}
(1993) 395]

\refer[[3]/G. 't Hooft, Nucl. Phys. {\bf B75} (1974) 461]

\refer[[4]/A.M. Polyakov and P.B. Wiegman, Phys. Lett. {\bf 131B} (1983) 121;
{\bf 141B} (1984) 223;]

\refer[/E. Witten, Commun. Math. Phys. {\bf 92} (1984) 455;]

\refer[/P. di Vecchia, B. Durhuus and J.L. Petersen, Phys. Lett. {\bf B144}
(1984) 245;]

\refer[/D. Gonzales and A.N. Redlich, Phys. Lett. {\bf B147} (1984) 150; Nucl.
Phys. {\bf B256} (1985) 621;]

\refer[/E. Abdalla and M.C.B. Abdalla, Nucl. Phys. {\bf B255} (1985) 392;]

\refer[/J.L. Petersen, Acta Phys. Polon. {\bf B16} (1985) 271]

\refer[[5]/V. Baluni, Phys. Lett. {\bf 90B} (1980) 407;]

\refer[/P.J. Steinhardt, Nucl. Phys. {\bf B176} (1980) 100;]

\refer[/D. Gepner, Nucl. Phys. {\bf B252} (1985) 481]

\refer[[6]/E. Abdalla and K. Rothe, Phys. Rev. {\bf D 361} (1987) 3190]

\refer[[7]/A. Patrasciou, Phys. Rev. {\bf D 15} (1977) 3592;]

\refer[/P. Mitra and P. Roy,  Phys. Lett. {\bf 79B} (1978) 469]

\refer[[8]/T.T. Wu, Phys. Rev. Lett. {\bf 71B} (1977) 142;]

\refer[/Y. Frishman, C.T. Sachrajda, H. Abarbanel and R. Blankenbecler,
Phys. Rev. {\bf D15} (1977) 2275;]

\refer[/E. Witten, Commun. Math. Phys. {\bf 141} (1991) 153]

\refer[[9]/A. Dhar, G. Mandal and S.R. Wadia, Phys. Lett. {\bf B329} (1994) 15]

\refer[[10]/E. Abdalla and M.C.B. Abdalla, CERN-TH 7354/94]

\refer[[11]/L. Alvarez-Gaum\'e, Trieste Summer School, Italy, 1993;]

\refer[/E. Alvarez, L. Alvarez-Gaum\'e and Y. Lozano, CERN-TH 7204/94,
hepth/9403155]

\refer[[12]/D. Karabali and H.J. Schnitzer, Nucl. Phys. {\bf B329} (1990) 649]

\refer[[13]/P. Goddard, A. Kent and D. Olive, Phys. Lett. {\bf B152} (1985)
88; Commun. Math. Phys. {\bf 103} (1986) 105.]

\refer[[14]/P.A.M. Dirac, ``Lectures on Quantum Mechanics", Yeshiva Univ.
Press,
N.Y. 1964; Can J. Math. {\bf 2} (1950) 129]

\refer[[15]/A. Cappelli, C.A. Trugenberger and  G.R. Zemba, Phys. Rev. Lett.
{\bf 72} (1994) 1902; Phys. Lett. {\bf B306} (1993) 100; Nucl. Phys. {\bf B396}
(1993) 465;]

\refer[/S. Iso, D. Karabali and B. Sakita, Phys. Lett. {\bf B296} (1992) 143]

\refer[[16]/V.G. Knizhnik and A.B. Zamolodchikov, Nucl. Phys. {\bf B247}
(1984) 83]

\refer[[17]/A.A. Belavin, A.M. Polyakov and A.B. Zamolodchikov, Nucl. Phys.
{\bf B241} (1994) 333]

\refer[[18]/E. Abdalla, M.C.B. Abdalla, J.C. Brunelli and A. Zadra,  to appear
in Commun. Math. Phys.]

\refer[[19]/E. Abdalla, M.C.B. Abdalla, G. Sotkov and M. Stanishkov, to appear
in Int. J. Mod. Phys. A]

\refer[[20]/A. Gorsky and N. Nekrasov, Nucl. Phys. {\bf B414} (1994) 213]

\refer[[21]/L.N. Lipatov, Phys. Lett. {\bf B251} (1990) 284; {\bf 309} (1993)
394; Sov. Phys. JETP {\bf 63} (1986) 904;]

\refer[/Ya. Balitsky and L.N. Lipatov, Sov. J. Nucl. Phys. {\bf 28} (1978)
822;]

\refer[/E. Verlinde and H. Verlinde, Princeton preprint PUPT-1319, Sept. 93]

\refer[[22]/L.D. Faddeev and G.P. Korchemsky, IPT-SB-94-14, hepth/9404173]

\refer[[23]/L.D. Faddeev, ``Lectures on quantum inverse scattering method" in
Nankai Lectures on Mathematical Physics and Integrable Systems, Ed. X.C. Song,
Singapore, World Scientific, 1990]

\refer[[24]/A. Jevicki, J.P. Rodrigues and A.J. van Tonder, Nucl. Phys.
{\bf B404} (1993) 91;]

\refer[/J. Avan and A. Jevicki, Mod. Phys. Lett. {\bf A7} (1992) 357; Nucl.
Phys. {\bf B 397} (1993) 672]
\end